\documentstyle[12pt]{article}

\begin{document}
\title{$\tau\rightarrow\eta(\eta')2\pi\nu,3\pi\nu$ and WZW anomaly}
\author{Bing An Li\\
Department of Physics and Astronomy, University of Kentucky\\
Lexington, KY 40506, USA}

\maketitle

\begin{abstract}
The effects of the anomalous contact terms ${\cal L}_{\eta(\eta')\rho
\pi\pi}$ are taken into account in calculating the decay rates of
$\tau\rightarrow\eta(\eta')\pi\pi\nu$. The branching ratio of
$\tau\rightarrow\eta\pi\pi\pi\nu$ is calculated. Theoretical result
agrees with the data. It is the first time that the anomalous
Wess-Zumino-Witten vertex ${\eta aa}$ is tested. A $a_{1}$ resonance
is predicted in the final state of the three pions.
The prediction of
the branching ratio of $\tau\rightarrow\eta'\pi\pi\pi\nu$ is presented.
\end{abstract}

\newpage
All the hadrons produced in $\tau$ hadronic decays are mesons
made of the light quarks, therefore,
the $\tau$ mesonic decays provide a test ground for all meson
theories.
Chiral symmetry plays an
essential role in studying $\tau$ mesonic decays[1].
The Wess-Zumino-Witten(WZW) anomalous action[2] is general and
model independent. It is
important part of the meson theory. Test of the WZW action is very
significant in the physics of strong interaction.
Various anomalous WZW vertices
have been used to calculate the decay widths of mesons[3,4].
The $\tau$ mesonic decays provide
a comprehensive test ground for WZW anomaly.
The abnormal vertices $\eta\rho\rho$ and $\omega\rho\pi$ derived from
the WZW anomalous action have been
studied[5] in $\tau$ mesonic decays. In Refs.[6] we have studied
more abnormal $\tau$ mesonic decays.
It is pointed out in Ref.[7] that the
$\eta$ production in $\tau$ decay is associated with anomaly.
Recently CLEO has reported a measurement of the branching ratio of
$\tau\rightarrow\eta(3h^{-})\nu$[8]
\begin{equation}
B(\tau^{-}\rightarrow\nu_{\tau}(3h^{-})\eta)=(4.1\pm 0.7\pm 0.7)\times
10^{-4}.
\end{equation}
In Ref.[7] an abnormal axial-vector
current of $\eta$ and pions has been constructed to calculate the
branching ratio of $\tau\rightarrow\eta3\pi\nu$.
The theoretical prediction is
\[1.2\times 10^{-6}\]
which is less than the experiment by more than two orders of magnitude.
The decay $\tau\rightarrow\eta3\pi\nu$ is caused by the axial-vector
current and, as pointed in Ref.[7], by the anomalous meson vertex.
The question is what anomalous meson vertices are responsible for this
decay mode? As shown in the studies[1,5,6], resonances play essential
role in $\tau$ mesonic decays.
In this letter we are
going to find all anomalous WZW terms which contribute to this decay
mode. In constructing the
anomalous vertices, the resonances $a_{1}$ and $\rho$ are needed to be
taken into account. In doing so, an effective chiral theory of mesons
is required.

In Ref.[4] an effective chiral theory of three nonets of
pseudoscalar, vector, and axial-vector mesons
is proposed. This theory has been used to study meson physics at low
energies. So far, it is phenomenologically successful.
The chiral symmetry breaking scale $\Lambda$ is
determined to be 1.6GeV[4], therefore, this theory is suitable
to be applied to study $\tau$ mesonic decays. Our studies of $\tau$
mesonic decays based on this effective theory
are presented in Ref.[6].
Theoretical results agree with the data reasonably well.
The Wess-Zumino-Witten action
is derived from the leading terms of
the imaginary part of the effective Lagrangian[4].
In this letter the Lagrangian of Ref.[4] is used to derive all vertices
needed.
The expression of the axial-vector current($\Delta s=0$)
of mesons presented in Ref.[6] is taken. The calculations done in this
letter are at the three level which is supported by the argument of
large $N_{C}$ expansion[4]. It is necessary to point out that in the
studies of this letter there is no any new parameter.

The anomalous vertices can be found in the WZW action[2,3].
The formalism of the WZW action is model independent. However,
the meson
fields in the WZW action need to be normalized to physical fields
by the normal part of the Lagrangian(see Refs.[3,4]).
In Ref.[4] a method is developed to derive the anomalous vertices which
are found to have the same expressions as the one's obtained from the
WZW action. In this way the meson fields are normalized to physical
mesons.
In this letter we use the method presented
in Ref.[4] to derive all anomalous vertices needed.

We first present the study of the new effect in $\tau\rightarrow\eta
\pi\pi\nu$ which
has been studied by many
author[5] in terms of the anomalous vertex $\eta\rho\rho$.
It is pointed out in this letter
that indeed, this vertex ${\cal L}_{\eta\rho\rho}$ plays major role in
the decay $\tau\rightarrow\eta\pi\pi\nu$, however,
there is another anomalous vertex which
contributes about $20\%$ of the decay rate. The study of
$\tau\rightarrow\eta'\pi\pi\nu$ is presented.
The anomalous WZW vertices derived in
Ref.[4] are
\begin{eqnarray}
{\cal L}_{\eta\rho\rho}=\frac{N_{C}}{2\pi^{2}g^{2}f_{\eta}}\varepsilon
^{\mu\nu\alpha\beta}\eta(-\sqrt{2\over3}sin\theta+{1\over\sqrt{3}}
cos\theta)\partial_{\mu}\rho^{i}_{\nu}\partial_{\alpha}\rho^{i}_{\beta},
\nonumber \\
{\cal L}_{\eta'\rho\rho}=\frac{N_{C}}{2\pi^{2}g^{2}f_{\eta'}}
\varepsilon
^{\mu\nu\alpha\beta}\eta'(\sqrt{2\over3}cos\theta+{1\over\sqrt{3}}
sin\theta)\partial_{\mu}\rho^{i}_{\nu}\partial_{\alpha}\rho^{i}_{\beta},
\end{eqnarray}
where $\theta=-20^{0}$ and \(f_{\eta}=f_{\eta'}=f_{\pi}\) are taken.
g is a universal coupling constant and is determined to be 0.39 in
Ref.[6].
Besides the vertices(2), there are other vertices which
contribute to these two decay modes. Using the same method
deriving Eqs.(2)[4], we obtain
\begin{eqnarray}
{\cal L}_{\eta\rho\pi\pi}=
\frac{2}{\pi^{2}gf^{3}_{\pi}}
(-\sqrt{2\over3}sin\theta+{1\over\sqrt{3}}
cos\theta)(1-{4c\over g}-{2c^{2}\over g^{2}})
\varepsilon^{\mu\nu\alpha\beta}
\epsilon_{ijk}\eta
\partial_{\mu}\rho^{i}_{\nu}\partial_{\alpha}\pi^{j}\partial
_{\beta}\pi^{k}
\nonumber \\
{\cal L}_{\eta'\rho\pi\pi}=
\frac{2}{\pi^{2}gf^{3}_{\pi}}
(\sqrt{2\over3}cos\theta+{1\over\sqrt{3}}
sin\theta)(1-{4c\over g}-{2c^{2}\over g^{2}})
\varepsilon^{\mu\nu\alpha\beta}
\epsilon_{ijk}\eta'
\partial_{\mu}\rho^{i}_{\nu}\partial_{\alpha}\pi^{j}\partial
_{\beta}\pi^{k}.
\end{eqnarray}
where \(c={f^{2}_{\pi}\over2gm^{2}_{\rho}}\).
The difference between this study and previous calculations[5]
is the inclusion of the anomalous contact terms(3) in calculating
the decay branching ratios.
In the final states of the decay $\tau\rightarrow\eta(\eta')\pi\pi\nu$
caused by the vertices(2) the two pions have a $\rho$-resonance
structure, while for the decay amplitudes determined by Eqs.(3)
there are no $\rho$
resonance. Because of the cancellation in the factor $1-{4c\over g}
-{2c^{2}\over g^{2}}$ the contributions of Eqs.(3) are smaller.
This is similar to the decay $\omega\rightarrow\pi\pi\pi$ which is
dominated by the vertex $\omega\rho\pi$ and the contribution of
the contact term $\omega\pi\pi\pi$ is small[3,4].
Using the VMD and Eqs.(2,3), the decay branching ratios are
calculated
\begin{eqnarray}
B(\tau\rightarrow\eta\pi\pi\nu)=1.9\times 10^{-3},\nonumber \\
B(\tau\rightarrow\eta'\pi\pi\nu)=0.44\times 10^{-5}.
\end{eqnarray}
The calculation shows that $28\%$ of $B(\tau\rightarrow\eta\pi\pi
\nu)$ is from the anomalous contact terms(3). The data are
\[B(\tau\rightarrow\eta\pi\pi\nu)=(1.71\pm 0.28)\times 10^{-3}[9],\]
\[B(\tau\rightarrow\eta'\pi\pi\nu)<0.8\times 10^{-4}[8].\]
Theoretical results agree with the data.

The decay process $\tau\rightarrow\eta\pi\pi\pi\nu$ is
more complicated than the one's studied above. Only
the axial-vector current
contributes to these processes. The WZW anomalous interactions of
mesons cause these decays. More vertices are involved.
The Feynman diagrams of these decays
are shown in Fig.1.
The derivation of these vertices is
lengthy. As done in Ref.[4], by
calculating $-{2i\over f_{\pi}}m\eta<\bar{\psi}\gamma_{5}\psi>$ and
$-{2i\over f_{\pi}}m\eta<\bar{\psi}\lambda_{8}\gamma_{5}\psi>$,
we derive all anomalous vertices contributing to
these decays:
\begin{eqnarray}
{\cal L}_{\eta aa}=\frac{f^{2}_{a}}{2\pi^{2}f_{\eta}}\varepsilon
^{\mu\nu\alpha\beta}\eta
(-\sqrt{2\over3}sin\theta+{1\over\sqrt{3}}cos\theta)
\partial_{\mu}a^{i}_{\nu}\partial_{\alpha}a^{i}_{\beta},
 \\
{\cal L}_{\eta\rho\pi\pi}={1\over4\pi^{2}g}({2\over f_{\pi}})^{3}
(1-{4c\over g}-{2c^{2}\over g^{2}})
(-\sqrt{2\over3}sin\theta+{1\over\sqrt{3}}cos\theta)
\epsilon_{ijk}\varepsilon^{\mu\nu\alpha\beta}\eta
\partial_{\mu}\rho^{i}_{\nu}\partial_{\alpha}\pi_{j}\partial_{\beta}
\pi_{k}, \\
{\cal L}_{\eta a\pi\pi}={N_{C}f_{a}\over
(4\pi)^{2}}({2\over f_{\pi}})^{4}
(-\sqrt{2\over3}sin\theta+{1\over\sqrt{3}}cos\theta)
(1-{10\over3}{c\over g}+{8\over3}{c^{2}\over g^{2}})
\varepsilon^{\mu\nu\alpha\beta}\eta\partial_{\mu}a^{i}_{\nu}
\partial_{\alpha}\pi^{2}\partial_{\beta}\pi^{i},
\end{eqnarray}
where $a^{i}_{\mu}$ is the $a_{1}$ meson field and \(f^{-1}_{a}=
g(1-{1\over2\pi^{2}g^{2}})^{{1\over2}}\).
The vertices related to $\eta'$ meson are obtained by using the
substitution
\[(-\sqrt{2\over3}sin\theta+{1\over\sqrt{3}}cos\theta)\eta
\rightarrow
(\sqrt{2\over3}cos\theta+{1\over\sqrt{3}}sin\theta)\eta'\]
in Eqs.(5-7).

The vertices ${\cal L}_{a\rho\pi}$ and ${\cal L}_{\rho\pi\pi}$ are
involved in $\tau\rightarrow\eta\pi\pi\pi\nu$. They are presented
in Refs.[4,6]
\begin{eqnarray}
\lefteqn{{\cal L}_{a_{1}\rho\pi}=\epsilon_{ijk}\{Aa^{i}_{\mu}
\rho^{j\mu}\pi^{k}-Ba^{i}_{\mu}\rho^{j}_{\nu}\partial^{\mu\nu}\pi^{k}\}
},\\
&&A={2\over f_{\pi}}gf_{a}\{{m^{2}_{a}\over g^{2}f^{2}_{a}}
-m^{2}_{\rho}+p^{2}[{2c\over g}+{3\over4
\pi^{2}g^{2}}(1-{2c\over g})]\nonumber \\
&&+q^{2}[{1\over 2\pi^{2}g^{2}}-
{2c\over g}-{3\over4\pi^{2}g^{2}}(1-{2c\over g})]\},\\
&&B=-{2\over f_{\pi}}gf_{a}{1\over2\pi^{2}g^{2}}(1-{2c\over g}),\\
&&{\cal L}^{\rho\pi\pi}={2\over g}\epsilon_{ijk}\rho^{i}_{\mu}
\pi^{j}\partial^{\mu}\pi^{k}-{2\over \pi^{2}f^{2}_{\pi}g}
\{(1-{2c\over g})^{2}-4\pi^{2}c^{2}\}\epsilon
_{ijk}\rho^{i}_{\mu}\partial_{\nu}\pi^{j}\partial^{\mu\nu}\pi^{k},
\end{eqnarray}
where p is the momentum of $\rho$ meson and q is
the momentum of $a_{1}$. These vertices have been used to calculate the
decay widths of $a_{1}\rightarrow\rho\pi$, $3\pi$, the ratio of
$d/s$, the width of $\rho\rightarrow\pi\pi$ and the width of
$\tau\rightarrow3\pi\nu$[4,6]. The theoretical results agree
with the data.

The axial-vector current is needed to calculate the decay rates
of $\tau\rightarrow\eta\pi\pi\pi\nu$.
In terms of chiral symmetry and dynamical chiral symmetry breaking,
the expression of the
axial-vector current of mesons is obtained in Ref.[6]
(Eqs.(10,23,24,25) of Ref.[6]). We use
this axial-vector current
to calculate some $\tau$ mesonic decay rates in Ref.[6]. Theoretical
results agree with data.
Using the axial-vector current and the vertices(Eqs.(5-7,8-11)),
the decay
rate of $\tau\rightarrow\eta\pi\pi\pi\nu$ is calculated.
In the subprocesses shown in Fig.1 there are anomalous vertices and
normal vertices. The anomalous vertices are at the forth order in low
energy expansion, hence the strengths of anomalous vertices are weaker.
This argument is supported by the narrow decay widths of the $\omega$
and $f_{1}(1280)$ mesons whose decays are resulted by anomalous
vertices[4]. Because the vertices(8,11) are normal and
at the second order
in low energy expansion, the strengths of the vertices(8,11)
are stronger.
Therefore, the $\rho$ and the $a_{1}$ mesons have broader widths.
Only the anomalous vertex ${\cal L}_{\eta a\pi\pi\pi}$ contributes to
the subprocess shown in Fig.1(a), therefore, the contribution of this
process to the decay is small. For the subprocess(Fig.1(b)) there are
normal vertices and anomalous ones.
For the anomalous vertex
${\cal L}_{\eta\rho\pi\pi}$(6),
because of the cancellation in the factor $(1-{4c\over g}
-{2c^{2}\over g^{2}})$,
the anomalous vertex
${\cal L}_{\eta\rho\pi\pi}$(6) is very weak. The contribution of this
subprocess is small. In the third subprocess(Fig.1(c)) there is a
$\rho$ resonance. The minimum of $q^{2}$ of this $\rho$
is \(m_{\rho}+m_{\eta}=1.32GeV\) which is much greater than $m_{\rho}$.
This effect suppresses the contribution of this subprocess.
As a matter of fact, the axial-vector current derived from the vertex
(7) is similar with the one presented in Ref.[7] in which it is shown
that the contribution of this vertex is very small.
Numerical calculation supports these arguments. The contribution
of these subprocesses is about few percents.
Finally, only the fourth subprocess(Fig.1(d)) survives.
There is no cancellation in the anomalous vertex ${\cal L}_{\eta aa}$
and the strength of the normal vertex ${\cal L}_{a\rho\pi}$ is strong.
Therefore, we expect that the diagram Fig.1(d) explains the decay rate
of $\tau\rightarrow\eta\pi\pi\pi\nu$.

The discussion above leads to that the process
$\tau\rightarrow\eta\rho\pi\nu$ is dominant
the decay $\tau\rightarrow\eta\pi\pi\pi\nu$. The Feynman diagram,
Fig.1(d), shows that there are two subprocesses in this decay:
$\tau\rightarrow\eta a_{1}\nu$ and $a_{1}\rightarrow\rho\pi$.
As mentioned above, the former is caused by the anomalous vertex
${\cal L}_{\eta aa}$(5) which is never tested before and
the later is resulted by the vertex(8) which
has been tested.
It is necessary to point out that the anomalous vertices
${\cal L}_{\eta vv}$(\(v=\rho, \omega, \phi\)) have been tested by
$\eta\rightarrow\gamma\gamma$, $\rho\rightarrow\eta\gamma$,
$\omega\rightarrow\eta\gamma$, $\phi\rightarrow\eta\gamma$, and $\tau
\rightarrow\eta\pi\pi\nu$. However, it is the first time that the
anomalous WZW vertex ${\cal L}_{\eta aa}$ is tested.

Using the vertices(5,8,11) and the axial-vector current of Ref.[6]
in the chiral limit
the amplitude of the decay $\tau\rightarrow\eta\rho\pi\nu$ is
derived as
\begin{eqnarray}
\lefteqn{<\eta(p)\rho^{0}(k')\pi^{-}(k)|\bar{\psi}\tau_{+}
\gamma_{\mu}\gamma_{5}\psi|0>
=\frac{i}{\sqrt{8\omega \omega'E}}\frac{f^{2}_{a}}{\pi^{2}f_{\pi}}
(-\sqrt{2\over3}sin\theta+{1\over\sqrt{3}}cos\theta)}\nonumber \\
&&\frac{g^{2}f_{a}m^{2}_{\rho}-if^{-1}_{a}\sqrt{q^{2}}\Gamma_{a}
(q^{2})}
{q^{2}-m^{2}_{a}+i\sqrt{q^{2}}\Gamma_{a}(q^{2})}
\frac{1}{q^{2}_{1}-m^{2}_{a}+i\sqrt{q^{2}_{1}}\Gamma_{a}
(q^{2}_{1})}
\{A(q^{2}_{1})g_{\beta\lambda}+Bk_{\beta}k_{\lambda}\}
\varepsilon^{\mu\nu\alpha\beta}q_{\nu}q_{1\alpha}
\epsilon^{*\lambda},
\end{eqnarray}
where \(q=p+k+k'\), \(q_{1}=q-p\), and k, k', and p are momentum of
pion, $\rho$, and $\eta$ respectively.
The width of the $a_{1}$ meson is derived from the vertex(8)
\begin{equation}
\Gamma_{a}(q^{2})={k_{a}\over12\pi}{1\over\sqrt{q^{2}}m_{a}}
\{(3+{k^{2}_{a}
\over m^{2}_{\rho}})A^{2}(q^{2})-2A(q^{2})B(q^{2}+m^{2})
{k^{2}_{a}\over m^{2}_{\rho}}+k^{4}_{a}{q^{2}\over m^{2}_{\rho}}B^{2}\},
\end{equation}
where \(k^{2}_{a}={1\over4q^{2}}(q^{2}+m^{2}_{\rho}-m^{2}_{\pi})^{2}
-m^{2}_{\rho}\).
Making the substitution $q^{2}\rightarrow q^{2}_{1}$ in Eq.(13), the
$\Gamma_{a}(q^{2}_{1})$ is obtained.

This matrix element has the strength of the vertex
${\cal L}_{a_{1}\rho\pi}$. In the matrix element(13) there is second
$a_{1}$ resonance
\[\frac{1}{q^{2}_{1}-m^{2}_{a}+i\sqrt{q^{2}_{1}}\Gamma_{a}
(q^{2}_{1})}.\]
The minimum of the $q^{2}_{1}$ is
$(m_{\rho}+m_{\pi})$ which is less than $m_{a}$. Therefore, the matrix
element is enhanced by the second $a_{1}$ resonance. These two factors
make the subprocess(Fig.1(d)) the main contributor of the decay
$\tau\rightarrow\eta\pi\pi\pi\nu$.
The term $g^{2}f_{a}m^{2}_{\rho}$ comes from dynamical chiral symmetry
breaking which is the origin of the mass difference of the $a_{1}$ and
the $\rho$ mesons[6].
Using the matrix element, in the chiral limit
the decay rate of $\tau\rightarrow\eta\rho
\pi\nu$ is found to be
\begin{eqnarray}
\lefteqn{d\Gamma(\tau^{-}\rightarrow\eta\rho^{0}\pi^{-}\nu)=
\frac{G^{2}}{(2\pi)^{5}}cos^{2}\theta_{c}{1\over1152}{1\over
m^{3}_{\tau}q^{2}}(m^{2}_{\tau}-q^{2})^{2}
(m^{2}_{\tau}+2q^{2})}\nonumber \\
&&(\frac{f^{2}_{a}a_{\eta}}{\pi^{2}f_{\pi}})^{2}\frac{m^{4}_{\rho}g^{2}
f^{2}_{a}+f^{-2}_{a}q^{2}\Gamma^{2}_{a}(q^{2})}{(q^{2}-m^{2}_{a})^{2}
+q^{2}\Gamma^{2}_{a}(q^{2})}\frac{1}{(q^{2}_{1}-m^{2}_{a})^{2}
+q^{2}_{1}\Gamma^{2}_{a}(q^{2}_{1})}\nonumber \\
&&\{3A^{2}(q^{2}_{1})\frac{(q\cdot p)^{2}}{q^{2}}(q^{2}_{max}-q^{2}
_{min})-{1\over6}A(q^{2}_{1})B(q\cdot p)^{2}{1\over q^{4}}
[(q^{2}-q^{2}_{min})^{3}-(q^{2}-q^{2}_{max})^{3}]\nonumber \\
&&+[A^{2}(q^{2}_{1})-2A(q^{2}_{1})Bk\cdot k'+B^{2}(k\cdot k')^{2}]\frac
{(q\cdot p)^{2}}{12m^{2}_{\rho}q^{4}}
[(q^{2}-q^{2}_{min})^{3}-(q^{2}-q^{2}_{max})^{3}]\}\nonumber \\
&&dq^{2}dq^{2}_{1},
\end{eqnarray}
where
\[q^{2}_{max}=m^{2}_{\rho}+{1\over\sqrt{q^{2}_{1}}}(q^{2}-q^{2}_{1})
(\sqrt{m^{2}_{\rho}+l^{2}}+l),\;\;\;
q^{2}_{min}=m^{2}_{\rho}+{1\over\sqrt{q^{2}_{1}}}(q^{2}-q^{2}_{1})
(\sqrt{m^{2}_{\rho}+l^{2}}-l),\]
where \(l={1\over 2\sqrt{q^{2}_{1}}}(q^{2}_{1}-m^{2}_{\rho})\) and \(
a_{\eta}=-\sqrt{2\over3}sin\theta+{1\over\sqrt{3}}cos\theta \).
The branching ratio is computed to be
\begin{equation}
B(\tau^{-}\rightarrow\eta\rho^{0}\pi^{-}\nu)=2.93\times 10^{-4}.
\end{equation}
Theoretical result agrees with the data reasonably well.
This theory predicts a $\rho$ resonance structure in the two pion
state and a $a_{1}$ resonance in the three pion state
of the decay $\tau\rightarrow\eta\pi\pi\pi\nu$.

Because of the value of $m_{\eta'}$ the decay $\tau\rightarrow
\eta'\rho\pi\nu$ is forbidden. Only the diagrams, Fig.1(a,b),
contribute to the decay. Because of kinematic reason
the $\rho$ resonance amplitude
in the amplitude derived from the vertex ${\cal L}_{\eta'\rho\pi\pi}$
provides a very strong suppression. The contribution of this
diagram is completely negligible. The decay rate derived from the
anomalous vertex ${\cal L}_{\eta' a\pi\pi\pi}$ is
\begin{eqnarray}
\lefteqn{d\Gamma(\tau^{-}\rightarrow\eta'\pi^{+}\pi^{-}\pi^{-}
\nu)=\frac{G^{2}cos^{2}\theta}{9216(2\pi)^{7}}\frac{p^{2}}{
48m^{3}_{\tau}q^{4}}F^{2}(m^{2}_{\tau}-q^{2})^{2}(m^{2}_{\tau}+2q^{2})
}\nonumber \\
&&[(q^{2}+p^{2}-m^{2}_{\eta'})^{2}-4q^{2}p^{2}]^{{1\over2}}
[{1\over4q^{2}}
(q^{2}+m^{2}_{\eta'}-p^{2})^{2}-m^{2}_{\eta'}][(q^{2}+p^{2}
-m^{2}_{\eta'})^{2}-q^{2}p^{2}]\nonumber \\
&&\frac{m^{4}_{\rho}g^{4}f^{4}_{a}+q^{2}\Gamma^{2}_{a}(q^{2})}
{(q^{2}-m^{2}_{a})^{2}+q^{2}\Gamma^{2}_{a}(q^{2})}dq^{2}dp^{2},\\
&&F=\frac{2N_{C}}{(4\pi)^{2}}({2\over f_{\pi}})^{4}(\sqrt{{2\over3}}
cos\theta+{1\over\sqrt{3}}sin\theta)(1-{10\over3}{c\over g}+{8\over3}
{c^{2}\over g^{2}}),
\end{eqnarray}
where \(q^{2}=(p_{\tau}-p_{\nu})^{2}\) and \(p^{2}=(q-p_{\eta'})^{2}
\),
\[B(\tau\rightarrow\eta'\pi\pi\pi\nu)=0.64\times 10^{-8}.\]

To conclude, it is predicted that
the anomalous contact terms of ${\cal L}_{\rho\eta\pi\pi}$ contributes
about $28\%$ of the decay rate of $\tau\rightarrow\eta\pi\pi\nu$.
The decay mode $\tau\rightarrow\eta\pi\pi\pi\nu$ provides the first
test on the anomalous vertex ${\cal L}_{\eta aa}$.
Theoretical results agree with data.
A $a_{1}$ resonance
structure in the final state of $\tau\rightarrow\eta\pi\pi\pi\nu$ is
predicted. The decay rate of $\tau\rightarrow\eta'\pi\pi\pi\nu$ is
predicted too. The smallness of the decay rate of $\tau\rightarrow\eta'
\pi\pi\pi\nu$ originates in the large value of $m_{\eta'}$ which is
not zero in the chiral limit.

The author wishes to thank E.Braaten
for suggesting this problem to the author
before the announcement of CLEO's
result[8].
This research was partially
supported by DOE Grant No. DE-91ER75661.

\newpage

\begin{center}
Fig.1 CAPTION
\end{center}
Diagrams of the decay $\tau\rightarrow\eta\pi\pi\pi\nu$
\end{document}